\documentclass[conference]{IEEEtran}
\IEEEoverridecommandlockouts

\usepackage{cite}
\usepackage{amsmath,amssymb,amsfonts}
\usepackage{algorithmic}
\usepackage{graphicx}
\usepackage{textcomp}
\usepackage{xcolor}
\usepackage{tabularx}
\usepackage{multirow}
\usepackage{makecell}
\usepackage{flushend}
\usepackage{hyperref}

\hypersetup{
    colorlinks=true,
    citecolor=blue,
    linkcolor=blue,
    filecolor=blue,      
    urlcolor=blue,
}

\def\BibTeX{{\rm B\kern-.05em{\sc i\kern-.025em b}\kern-.08em
    T\kern-.1667em\lower.7ex\hbox{E}\kern-.125emX}}

\begin{document}

\title{SRNI-CAR: A Comprehensive Dataset for Analyzing the Chinese Automotive Market}

\makeatletter
\newcommand{\linebreakand}{%
  \end{@IEEEauthorhalign}
  \hfill\mbox{}\par
  \mbox{}\hfill\begin{@IEEEauthorhalign}
}
\makeatother

\author{Ruixin Ding\\
\IEEEauthorblockA{
\textit{Adam Smith Business School} \\
\textit{University of Glasgow}\\
2589583d@student.gla.ac.uk}
\and
Bowei Chen\\
\IEEEauthorblockA{
\textit{Adam Smith Business School} \\
\textit{University of Glasgow}\\
bowei.chen@glasgow.ac.uk}
\and
James M. Wilson\\
\IEEEauthorblockA{
\textit{Adam Smith Business School} \\
\textit{University of Glasgow}\\
james.wilson@glasgow.ac.uk}
\linebreakand
Zhi Yan\\
\IEEEauthorblockA{
\textit{CIAD UMR 7533} \\
\textit{UTBM}\\
zhi.yan@utbm.fr}
\and
Yufei Huang\\
\IEEEauthorblockA{
\textit{Trinity Business School} \\
\textit{Trinity College Dublin}\\
yufei.huang@tcd.ie}
}

\maketitle

\begin{abstract}
The automotive industry plays a critical role in the global economy, and particularly important is the expanding Chinese automobile market due to its immense scale and influence. However, existing automotive sector datasets are limited in their coverage, failing to adequately consider the growing demand for more and diverse variables. This paper aims to bridge this data gap by introducing a comprehensive dataset spanning the years from 2016 to 2022, encompassing sales data, online reviews, and a wealth of information related to the Chinese automotive industry. This dataset serves as a valuable resource, significantly expanding the available data. Its impact extends to various dimensions, including improving forecasting accuracy, expanding the scope of business applications, informing policy development and regulation, and advancing academic research within the automotive sector. To illustrate the dataset's potential applications in both business and academic contexts, we present two application examples. Our developed dataset enhances our understanding of the Chinese automotive market and offers a valuable tool for researchers, policymakers, and industry stakeholders worldwide.
\end{abstract}


\section{Introduction}

The automotive industry has been one of the important economic sectors in many countries and regions. It also has an important role in global energy consumption, energy sustainability and environmental impact. At present, China has become the world's largest market for new energy vehicles (NEVs) and one of the largest automobile markets. Studying the Chinese automobile market reveals its profound impact on the global automotive ecosystem in terms of market growth opportunities, innovation cooperation, market diversity, environmental sustainability, and more. For a long time, the automotive industry has been facing many challenges in the marketing process, such as fierce market competition, the diversity of consumer demand, and the pressure of environmental protection policies. With the growth of social media and digital advertising, consumer engagement on the Web has greatly increased, and rational and effective digital marketing has become another serious challenge in the automotive industry~\cite{homburg2022value}.

The imperative for precise and efficient forecasting, rooted in historical data, becomes evident to address the challenges mentioned above. A substantial cohort of automotive industry stakeholders including executives, marketers, and academics are actively in pursuit of more efficacious methods for market analysis. Traditional forecasting approaches face formidable hurdles when contending with vast datasets, often falling short in ensuring the fidelity of sales predictions~\cite{cheriyan2018intelligent}. While the advent of cutting-edge data-driven methods such as data mining and machine learning has led to significant advancements in knowledge discovery and predictive capabilities, their efficacy hinges crucially on the availability of comprehensive and accurate datasets. However, currently available datasets are often fragmented, making it difficult for researchers to integrate industry information, automakers' behavior, consumer demand, market feedback, and sales forecasts. It is increasingly important to create a dataset that can meet the multiple business analytics needs of the automotive industry. Therefore, we collected car model sales data, consumer online review data, and automotive industry news and information data from different sources, and integrated them into a comprehensive dataset, called \emph{SRNI-CAR}.

\begin{table*}[htp]
\centering
\caption{Summary of variables in the datasets for automotive market or consumer analytics.}
\label{RelatedData}
\scalebox{0.96}{
\begin{tabular}{l|cccccc|c}
\hline
Variable &~\cite{xia2020forexgboost}  &~\cite{geva2017using}  &~\cite{chen2011role}  &~\cite{landwehr2011gut}  &~\cite{sa2012multi} &~\cite{wang2018dynamic} & Our dataset \\ \hline
Vehicle attribute related variables (e.g., model type, model size, energy type, etc.)  & \checkmark        &          & \checkmark        & \checkmark        & \checkmark        &          & \checkmark                                                              \\
Brand related variables (e.g., brand energy type, brand country-of-origin, etc.)    & \checkmark        & \checkmark        &          & \checkmark        &          &          & \checkmark                                                              \\
Price variables (e.g., official price, transaction price, etc.)              & \checkmark        &          & \checkmark        & \checkmark        &          & \checkmark        & \checkmark                                                              \\
Date-related variables (e.g., year of purchase, year of review, model launch date, etc.)   & \checkmark        & \checkmark        & \checkmark        & \checkmark        & \checkmark        & \checkmark        & \checkmark                                                              \\
Consumer experience variables (e.g., experience duration, mileage, etc.)&          &          &          &          &          & \checkmark        & \checkmark                                                              \\
User ratings (e.g., overall rating, exterior rating, interior rating, etc.)&          &          & \checkmark        &          &          & \checkmark        & \checkmark                                                              \\
Online reviews (e.g., advantage, features comments, comfort comments, etc.)        &          & \checkmark        & \checkmark        &          &          & \checkmark        & \checkmark                                                              \\
Industry news and information (e.g., title, text, information label, pageview, etc.) &          &          &          &          &          &          &\checkmark                                                              \\
Sales                                                                              & \checkmark        & \checkmark        & \checkmark        & \checkmark        & \checkmark        &          & \checkmark                                                              \\
Website search variables (e.g., search volume of brand, etc.) &          & \checkmark        &          &          &          &          &                                                                \\
Economic factor (e.g., consumer price index, gas prices, etc.)    &          & &          &          & \checkmark        &          &                                                                \\
Vehicle design (e.g., vehicle pictures, design fluency, etc.)     &          &          &          & \checkmark        &          &          &                                                                \\ \hline
\end{tabular}}
\end{table*}

Our work makes two significant contributions. Firstly, we have curated a more comprehensive and scalable data resource. It not only consolidates industry news, development insights, automotive marketing data, consumer online reviews, and sales information but also introduces valuable variables previously absent, such as model launch dates and brand inception dates. Therefore, our dataset supports a broader spectrum of research possibilities compared to existing publicly available datasets in the automotive domain. Furthermore, it enhances analytical accuracy and interpretability. Secondly, our dataset possesses substantial business value in the automotive sector. Sales data aids automakers and marketers in discerning market trends, while review data facilitates the identification of consumer preferences, evaluation of marketing effectiveness, and product strengths and weaknesses analysis. Incorporating industry news, development insights, and automotive publicity empowers automakers to grasp industry trends and market competition. Integrated analysis across multiple data provides superior decision support for product planning, business expansion, and marketing strategy formulation. Our dataset can be used to improve product quality, and align with consumer demands. Additionally, our dataset can be used by government policymakers, enabling the formulation of pertinent automotive industry policies and regulations. Thus, the development of this dataset holds profound significance for a myriad of stakeholders within the automotive industry.

\section{Review of Existing Automotive Datasets}

The automobile industry has garnered significant attention from researchers and analysts, particularly for sales forecasting. A review of prior studies reveals that researchers commonly incorporate a range of variables encompassing vehicle attributes, brand characteristics, pricing, temporal factors, user experiences, ratings, reviews, website search data, economic indicators, and vehicle design to prognosticate car sales. However, as shown in Table~\ref{RelatedData}, none of the datasets employed in these studies encompass all these pertinent variables. In the realm of nonlinear modeling, the reduction of variables often proves counterproductive to enhancing predictive model performance~\cite{hulsmann2012general}. Consequently, the absence of comprehensive variables constitutes a notable deficiency within the existing datasets in the automotive industry.

The limitations of the current automotive datasets are noteworthy. Firstly, they often omit vital information about brand creation and model launch dates, which is crucial for understanding market dynamics~\cite{robinson1985sources}. Secondly, they fail to differentiate between the growing number of new energy vehicle brands and do not clearly identify brand origins, despite their influence on consumer perceptions ~\cite{haubl1996cross}. Thirdly, these datasets often lack detailed comments and ratings for specific vehicle attributes, limiting their usefulness for sales forecasting and preference analysis ~\cite{geva2017using}. Fourthly, they usually provide only aggregate pricing data for vehicle series, omitting separate pricing for individual models and making it challenging to analyze the impact of discounts ~\cite{wang2018dynamic}. Furthermore, important data related to the automotive industry, such as model sentiment and review articles, are often missing, and aligning them with sales data remains challenging ~\cite{urban1990prelaunch}. These limitations highlight the need for more comprehensive datasets encompassing critical temporal, attribute-specific, and contextual factors for robust automotive market analysis and forecasting.

\section{Data Collection, Preparation and Description}

We collected sales and online review data from PCAuto and Dongchedi, and automotive industry news and information data from Autohome. They are three most authoritative and largest automobile media platforms in China. 

While the collected raw data encompass a broader range of variables than previous studies, they still lack essential elements necessary to address the evolving requirements of business analysis and academic research. To address these gaps, we introduced additional variables related to brand country-of-origin classification, vehicle entry order, and brand entry order. We also synchronized monthly sales data with online reviews to enhance our ability to derive valuable business insights. Furthermore, we manually incorporated actual transaction prices and official guide prices for specific car models within each series to investigate pricing and discount impacts. To ensure data validity, we diligently identified and addressed missing and outlier values across three data sources. Sales data required minimal processing, containing no missing values, outliers, or duplicates. For online reviews, we selectively removed instances with multiple missing values and those with missing values in less than 1\% of variables other than sales. Notably, data with only a missing value in sales were retained, comprising 9.68\% of the total dataset. It is important to note that imported car sales data are not included, resulting in the absence of corresponding sales data in the review dataset. Finally, we meticulously eliminated duplicates to produce the refined dataset intended for publication.

\begin{table*}[htp]
\centering
\caption{Description of variables in SRNI-CAR.}
\label{variable}
\begin{tabular}{p{1.25cm}|p{3.55cm}|p{12cm}}
\hline
\textbf{Data}  & \textbf{Variables}       & \textbf{Description}\\ 
\hline
\multirow{12}{*}{Sales}    & Car series         & Name of the car series.  \\
& Brand                    & Name of the brand.    \\
& Year                     & Year in which the car series was sold.\\
& Month                    & Month in which the car series was sold. \\
& Car model type           & Car model category: Sedan, SUV and MPV. \\
& Brand energy type        & Brand category based on energy type of vehicle produced.\\ 
& Size                     & Vehicle size category: mini, minivan, minibus, small, compact, mid-size, larger than mid-size, full-size.  \\
& Brand country of origin  & Country in which the brand was created.  \\
& Model launch date        & Year when the car series was launched on the Chinese market.  \\
& Brand establishment date & Year when the brand was created.  \\
& Brand entered China date & Year when the brand officially entered the Chinese market. \\
& Sales                    & Total sales of the car series in the month.\\ 
\hline
\multirow{45}{*}{Reviews}     & Car series  & Name of the car series. \\
& Brand                       & Name of the brand. \\
& Size                        & Vehicle size category: mini, minivan, small, compact, mid-size, larger than mid-size, full-size. \\
& Car model type              & Car model category: Sedan, SUV and MPV. \\
& User ID                     & Name that users use when making online reviews. \\
& Year of review              & Year when the user reviews. \\
& Month of review             & Month when the user reviews. \\
& Specific model purchased    & Specific model of a car series purchased by a user. \\
& Official price              & Official prices for specific models purchased. \\
& Car energy type             & Vehicle energy type: gasoline vehicle, diesel vehicle, hydrogen vehicle, and so on.  \\
& Brand energy type           & Brand category based on energy type of vehicle produced. \\ 
& Brand country of origin     & Country in which the brand was created. \\
& Brand establishment date    & Year when the brand was created. \\
& Brand entered China date    & Year when the brand officially entered the Chinese market. \\
& Model launch date           & Year when the car series was officially launched on the Chinese market. \\
& Year of purchase            & Year when the user purchased the model. \\
& Month of purchase           & Month when the user purchased the model. \\
& Sales                       & Total sales of the car series in the month when the user purchased the model. \\
& Experience duration         & Months between purchase date and review posting date.  \\
& Province                    & Province in which the user purchased the model.  \\
& City                        & City in which the user purchased the model. \\
& Transaction price           & Real transaction price of the model. \\
& Average energy consumption  & Gasoline, diesel, electricity, or hydrogen consumed for every 100 kilometers traveled. \\
& Mileage                     & Kilometers the user has driven the model at the date the review was posted. \\
& Overall rating              & User's overall rating of the vehicle purchased. \\
& Exterior rating             & User's rating of the exterior of the vehicle. \\
& Interior rating             & User's rating of the interior of the vehicle.  \\
& Space rating                & User's rating of the space of the vehicle. \\
& Features rating             & User's rating of the feature of the vehicle.  \\
& Power rating                & User's rating of the power of the vehicle.  \\
& Energy consumption rating   & User's rating of the energy consumption of the vehicle.  \\
& Driving rating              & User's rating of the driving of the vehicle.  \\
& Comfort rating              & User's rating of the comfort of the vehicle. \\
& Advantage                   & The advantages of the model as perceived by the user. \\
& Disadvantage                & The disadvantage of the model as perceived by the user. \\
& Exterior comments           & User's comments on the exterior of the vehicle.  \\
& Interior comments           & User's comments on the interior of the vehicle.  \\
& Space comments              & User's comments on the space of the vehicle. \\
& Features comments           & User's comments on the feature of the vehicle. \\
& Power comments              & User's comments on the power of the vehicle. \\
& Energy consumption comments & User's comments on the energy consumption of the vehicle. \\
& Driving comments            & User's comments on the driving of the vehicle. \\
& Comfort comments            & User's comments on the comfort of the vehicle. \\ 
\hline
\multirow{10}{*}{\begin{tabular}{@{}c@{}}News\\ information\end{tabular}} & Title  & Title of the information. \\
& Pageview           & Number of times the information was viewed.                                                       \\
& Number of comments & Number of comments the information received.                                                      \\
& Text               & Text content contained in the information.                                                        \\
& Release date       & Date on which this information was published.                                                     \\
& Author             & Person who posted the information.                                                                \\
& Source             & Source of the information.                                                                        \\
& Information type   & Whether the information is original, compiled, a press release, or reprinted from another platform. \\
& Information label  & Labels chosen by the author that summarizes the information, based on its content.                   \\ 
\hline
\end{tabular}
\end{table*}


As shown in Table~\ref{variable}, SRNI-CAR consist of sales, online reviews, and automotive industry news and information, over the period from 2016 to 2022. The dataset can be downloaded from the following address:
\begin{center}
\href{https://srni-car.github.io}{https://srni-car.github.io}
\end{center}

The sales data comprises 1,236 car series, including 518 sedans, 598 SUVs, and 121 MPVs, with 39,496 observations. It encompasses 155 car brands, with 107 traditional and 48 new energy brands, originating primarily from Germany, France, Korea, the Czech Republic, the USA, Japan, Sweden, Italy, the UK, and China. This data is valuable for research in sales forecasting, first-mover advantages, brand country-of-origin, and consumer preferences. It's stored as a 3.6 MB CSV file. The online review data is based on 217,292 comments from car owners in 358 cities, covering 13,039 specific models across 672 car series, ranging from 26,800 to 14.88 million yuan. It includes 10,977 traditional and 2,062 new energy models under 127 car brands, categorized into eight aspects by vehicle attributes. It also contains official and real transaction prices for precise analysis of discounts' impact, which can support research in sales forecasting, consumer behavior, product evaluation, and fake review detection, stored in a 480 MB CSV file. The automotive industry news and information data contains 83,590 items that covers a wide range of topics. Each data items is accurately labeled, aiding researchers in selecting relevant data. This data is stored as a 224.1 MB CSV file, beneficial for analyzing industry trends, marketing strategies, consumer perceptions, and automotive technology development.

\section{Automotive Analytics Examples}

Two application examples are presented to showcase the potential of our dataset in automotive analytics.

\subsection{Automobile Sales Forecasting}

For automotive manufacturers, proficiency in sales forecasting holds critical significance in shaping both product development and marketing strategies. The intricate interplay among an array of factors influencing sales, including price dynamics\cite{assuncao1993rational, busse2010best}, strategic discounting, consumer sentiment gleaned from online reviews, and numerical ratings~\cite{hu2014ratings}, necessitates a comprehensive analytical framework. Moreover, analysts must diligently consider factors like first-mover advantages, brand country-of-origin (COO), brand attributes, product attributes, and the increasingly salient energy characteristics of vehicles, especially within the context of China's fervent pursuit of NEVs~\cite{zahoor2023can}. Unlike previous studies constrained by data limitations, our dataset offers an opportunity for an extensive and all-encompassing examination of these pivotal determinants of sales.

\begin{figure*}[htp]
\begin{minipage}[t]{0.5\textwidth}
\centering
\includegraphics[height=22cm]{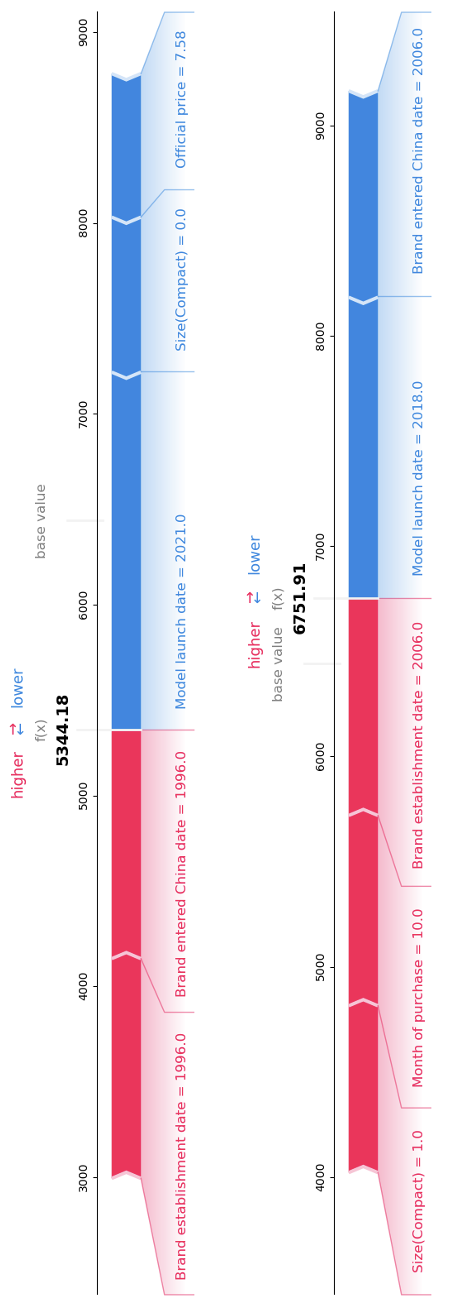}
\caption{Top variables with the most significant contributions in two data instances.}
\label{Example2}
\end{minipage}%
\begin{minipage}[t]{0.5\textwidth}
\centering
\includegraphics[height=22cm]{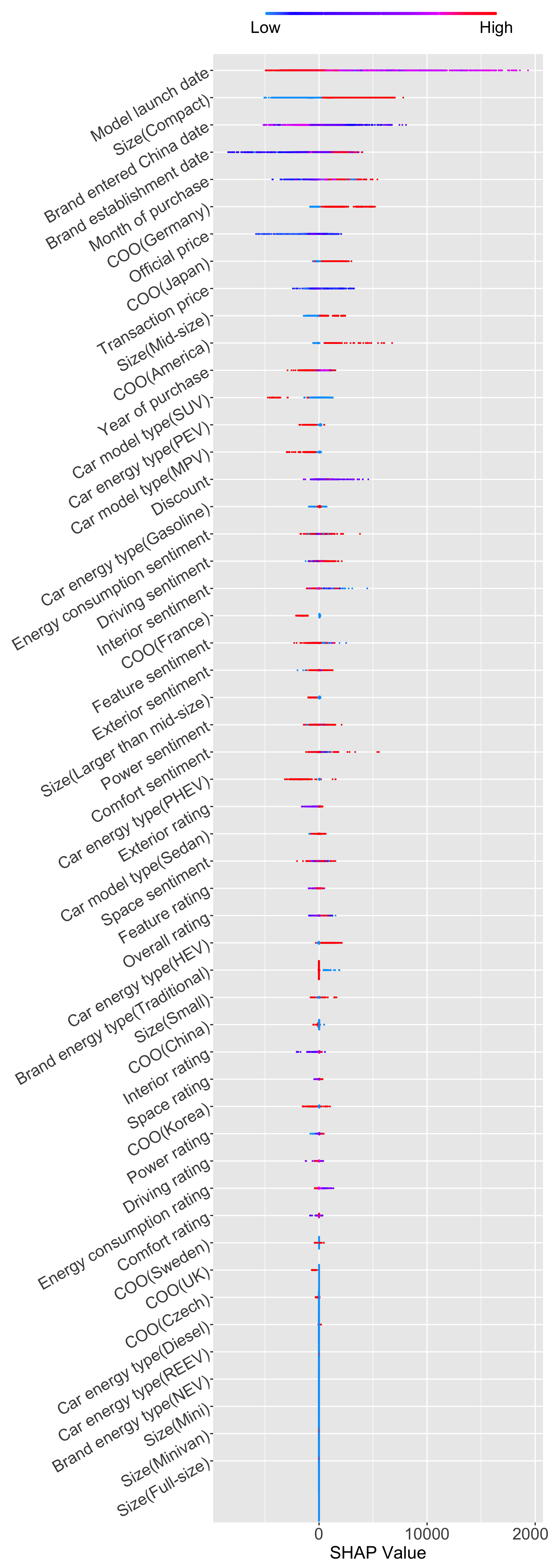}
\caption{Shap values corresponding to the variables influencing car sales.}
\label{SHAP}
\end{minipage}
\end{figure*}

\begin{figure*}[htp]
\centering
\includegraphics[width=1\textwidth]{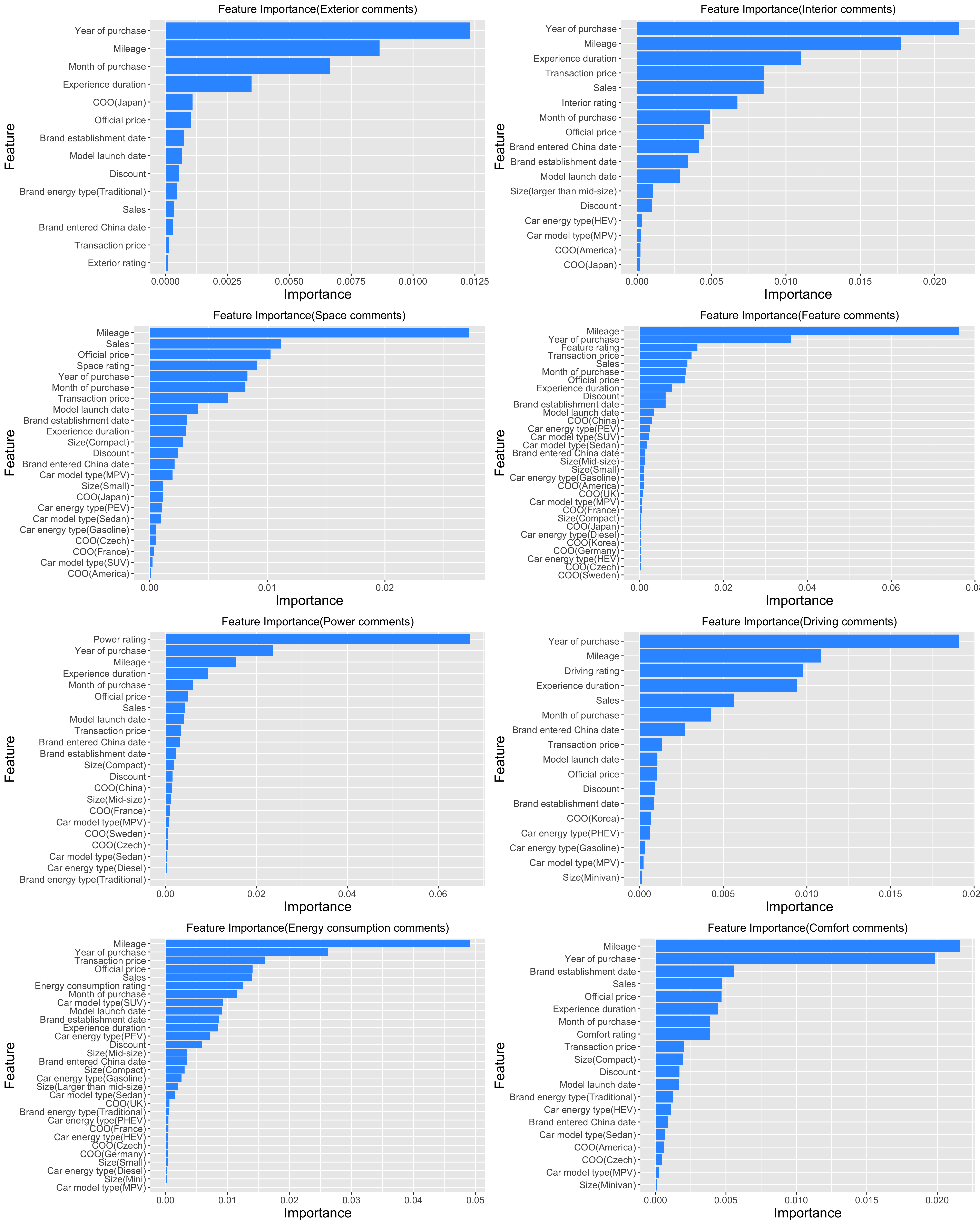}
\caption{Importance of variables in predicting sentiment across eight review categories.}
\label{ReviewImportance}
\end{figure*}

\begin{figure*}[htp]
\centering
\includegraphics[width=0.63\linewidth]{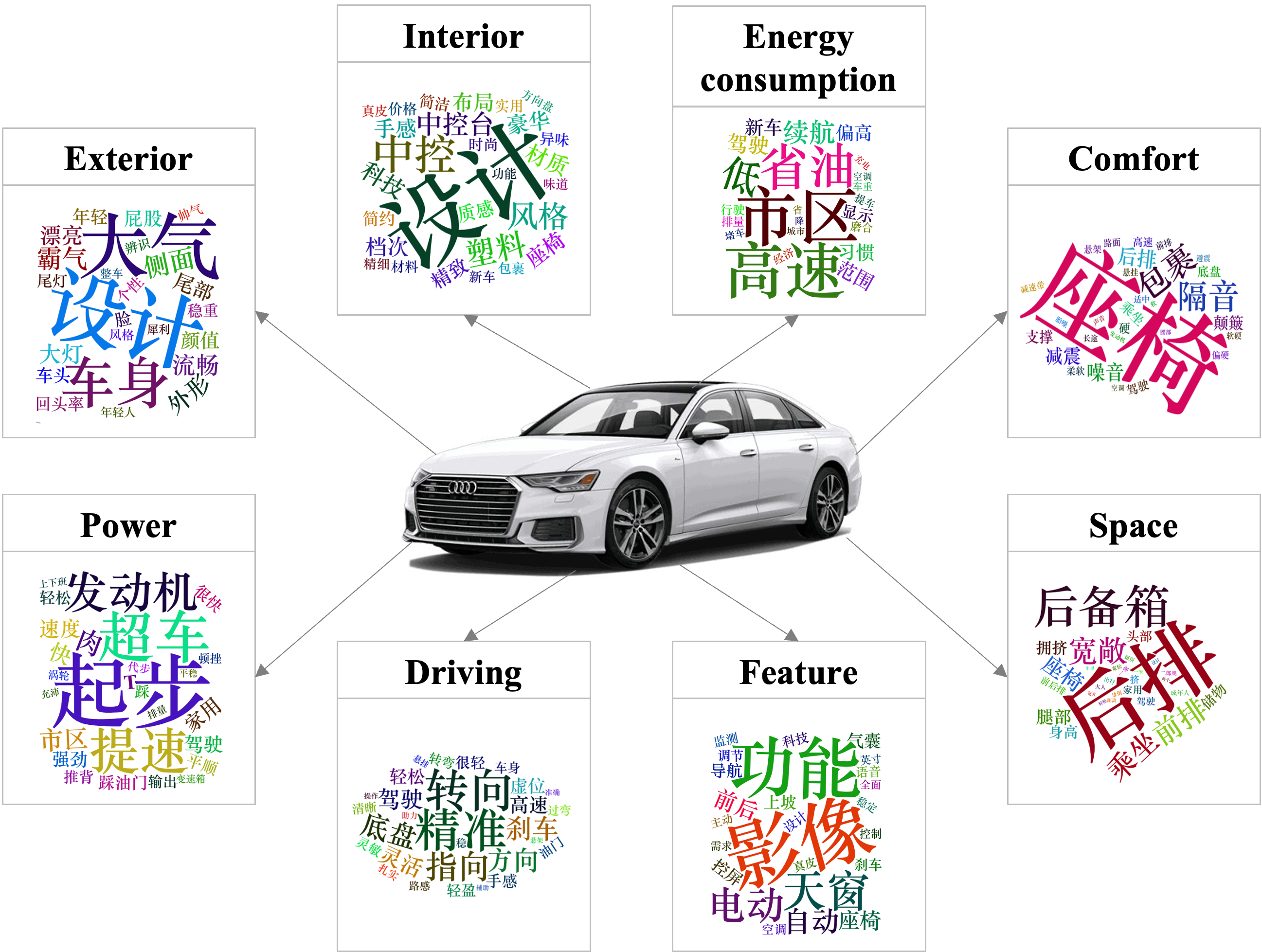}
\caption{Word clouds created from review comments associated with the eight vehicle attributes.}
\label{wordcloud}
\end{figure*}

To assess the impact of these factors on sales, we harnessed online review data spanning 2019 to 2021 and opted for the XGBoost model, lauded for its efficiency and precision in sales forecasting~\cite{shilong2021machine}. We optimized model parameters using grid search cross-validation to bolster predictive performance. To add robustness, one can explore alternative machine learning methodologies but that is not the most significance of our example here. Categorical predictors were judiciously transformed into dummy variables. Furthermore, we employed SnowNLP for sentiment analysis of textual comments, yielding sentiment scores on a scale from 0 to 1 as supplementary predictive variables. Finally, to interpret XGBoost model outputs, we applied the Shapley additive explanations (in short SHAP), mainly for its efficacy in elucidating feature contributions without the encumbrance of multi-collinearity concerns~\cite{marcilio2020explanations, baptista2022relation}.

In our analysis, we were able to obtain localized insights, presented in the form of Fig.~\ref{Example2}, which depict the contribution of each variable for every data instance—commonly referred to as Shap values. In a locally interpreted force plot, we can determine the contribution of features in this data instance by observing the length of the arrows. Additionally, red arrows represent features that positively impact the prediction, while blue arrows indicate a negative impact. For example, in the first data instance, we can observe that model launch date has the largest contribution to the prediction, but is negatively impacting it. On the other hand, brand entered China date, although its contribution is relatively lower, still positively impacts the prediction. Therefore, this data instance reveals that the launch date of the car model is a significant factor influencing its sales, albeit with a negative effect. Simultaneously, brand entered China date is also an important factor, despite its relatively lower contribution, as it positively affects the sales. Within the SHAP framework, variable importance is gauged as the average of absolute Shap values across all data instances for each variable. Fig.~\ref{SHAP} comprehensively outlines the variables employed in our experiment, showcases the experimental outcomes, and ranks these variables based on their significance in predicting sales. Notably, it visually underscores how the contribution of each variable to sales evolves as the variable's value changes.

We found several intriguing and valuable findings. Firstly, the sequence of model and brand entry wields a significantly higher influence on sales compared to other variables. Notably, alongside vehicle size, the launch date of a model, brand entry date into China, and the brand's founding date emerge as the three most critical factors. This implies the existence of both model and brand first-mover advantages in the Chinese automobile market, albeit with the caveat that late-mover advantages are still relevant based on the brand's entry date into China. Secondly, within the realm of price-related variables, although discounts exert a promotional effect on sales, official prices and transaction prices exert more substantial impacts, with official prices retaining the highest importance. However, it is worth noting that the relationship between price and sales is not linear. Thirdly, sentiment extracted from review text, categorized by vehicle attributes, exerts a notably higher impact on sales compared to corresponding rating data. This suggests that the review text serves as a more representative source of consumer sentiment regarding a model. Fourthly, within the Chinese automotive market, compact models and those from German, Japanese, and American brands maintain their popularity among consumers. Lastly, concerning new energy vehicles and associated brands, consumers exhibit a preference for pure electric vehicles (PEVs) and hybrid electric vehicles (HEVs), alongside a high level of acceptance for NEV brands. 

In summary, these findings underscore the imperative for automaker managers and marketers to meticulously analyze and weigh these factors when devising enterprise development and marketing strategies. Such insights are also instrumental for government departments in evaluating the effectiveness of new energy vehicle policies.

\subsection{Consumer Behavior Analytics}

Online reviews, generated by consumer post-purchase, significantly reflect and influence consumer decision-making~\cite{chen2008online}. Rating and review text are key components of online reviews, with ratings indirectly affecting sales through their interaction with the text~\cite{li2019effect}. Consumers tend to use ratings to shortlist products and rely on review text for final choices~\cite{hu2014ratings}, highlighting the higher business value of review text in expressing customer satisfaction. Understanding how consumer behavior and preferences impact review sentiment is crucial for shaping marketing strategies.  

Previous studies have shown that the level of user experience and price have an impact on automotive customer satisfaction. Mileage and experience duration are two important variables in the automotive industry to measure the level of experience~\cite{wang2018dynamic}. In addition, because consumers may have different expectations for different countries and different types of products, factors such as brand country-of-origin, brand history, and car type may also have an impact on the sentiment of the review text. At the same time, previous studies did not take into account that the factors that affect the sentiment of reviews for different attributes of vehicles may be different. For example, perhaps the review for comfort is more likely to be influenced by the level of user experience than the review of exterior. SRNI-CAR makes it possible to forecast the sentiment of automobile consumer reviews and to study the above questions.

In this example, we conducted sentiment score forecasts for comments associated with distinct vehicle attributes, employing the SnowNLP, XGBoost, and SHAP. The comprehensive experiment was divided into eight segments, each aligned with specific vehicle attributes. These segments encompassed the forecasting of sentiment scores for comments pertaining to exterior, interior, space, features, power, driving, energy consumption, and comfort attributes.

The depicted analysis Fig.~\ref{ReviewImportance} underscores the significance of various variables in predicting sentiment scores for comments linked to diverse vehicle attributes. The combination of these eight forecasts yields several pivotal insights. First, it is evident that the user's experience level, encompassing driving mileage and experience duration, influences consumer review sentiments, indirectly affirming its impact on overall customer satisfaction, with mileage proving more influential than experience duration among the eight vehicle attributes examined. Second, price-related factors indeed influence consumer reviews, with transaction and official prices holding greater sway than discounts, although their influence is less pronounced in reviews concerning comfort, exterior, and power attributes. Third, the year of vehicle purchase emerges as a crucial factor in forecasting review sentiment, an aspect hitherto unexplored, suggesting evolving consumer expectations. Fourth, vehicle-related attributes, such as size and energy type, demonstrate little relevance in exterior-related comments, indicating that Chinese consumers lack a distinct preference for specific model appearances. Fifth, model launch date, brand establishment date, and entry into the Chinese market date consistently wield importance across all eight review categories, suggesting that consumer review sentiment may signify either a first-mover or late-mover advantage. Sixth, ratings assume the highest significance in reviews associated with power, with elevated importance in feature, driving, and space-related reviews, highlighting the divergence between ratings and textual comments across various review aspects. Last but not least, the country of origin, size, and energy type exhibit varying impacts on sentiment in comments across different aspects. Furthermore, as shown in Fig.~\ref{wordcloud}, we conducted a word frequency analysis of comments pertaining to the eight distinct attributes and subsequently generated word clouds. The findings elucidate consumers' distinct priorities within these attributes. Notably, consumers focus more on the exterior design when assessing a vehicle's appearance. Concerning interior design, particular attention is directed toward the vehicle's central console. In evaluating power performance, consumers prioritize acceleration capabilities, while precision in steering takes precedence in assessments of handling. Features such as the reverse camera and functionality garner significant attention. The spaciousness of rear-seat areas is a key consideration for consumers when assessing vehicle space. In the realm of comfort, the quality of seats emerges as a pivotal factor. Lastly, regarding energy consumption, consumers exhibit heightened scrutiny when driving in urban settings and on highways. 

In summary, when exploring consumer behavior within the automobile market, one must consider the multifaceted nature of review data. Our dataset and accompanying analysis provide invaluable insights for marketers, facilitating the evaluation of product competitiveness, the identification of suitable marketing strategies, and the implementation of precision marketing tactics to enhance overall marketing effectiveness.

\section{Conclusion}

We have diligently created a comprehensive dataset SRNI-CAR for Chinese automotive market. This paper details the data collection and processing procedures, expounds upon the intricacies of variables, and highlights the dataset's profound relevance in areas such as market demand, consumer behavior, preferences, and industry development. We believe that SRNI-CAR will benefit not only the academic research community but also automotive industry leaders, marketers, and government agencies. We remain committed to continuously improving this dataset, with plans for updates and expansions, all aimed at enhancing its practical applications. This paper serves as a means to make the dataset accessible to a wider audience, facilitating a deeper understanding of the complexities within the automotive sector.

\section*{Acknowledgment}
The first author acknowledges the University of Glasgow Adam Smith Business School for Student Research Internship funding support. The second author would like to thank the funding support of Nvidia Accelerated Data Science Grant and University of Glasgow Adam Smith Business School.

\bibliographystyle{IEEEtran}
\bibliography{IEEEfull,mybibfile}

\begin{thebibliography}{10}
\providecommand{\url}[1]{#1}
\csname url@samestyle\endcsname
\providecommand{\newblock}{\relax}
\providecommand{\bibinfo}[2]{#2}
\providecommand{\BIBentrySTDinterwordspacing}{\spaceskip=0pt\relax}
\providecommand{\BIBentryALTinterwordstretchfactor}{4}
\providecommand{\BIBentryALTinterwordspacing}{\spaceskip=\fontdimen2\font plus
\BIBentryALTinterwordstretchfactor\fontdimen3\font minus \fontdimen4\font\relax}
\providecommand{\BIBforeignlanguage}[2]{{%
\expandafter\ifx\csname l@#1\endcsname\relax
\typeout{** WARNING: IEEEtran.bst: No hyphenation pattern has been}%
\typeout{** loaded for the language `#1'. Using the pattern for}%
\typeout{** the default language instead.}%
\else
\language=\csname l@#1\endcsname
\fi
#2}}
\providecommand{\BIBdecl}{\relax}
\BIBdecl

\bibitem{homburg2022value}
C.~Homburg and D.~M. Wielgos, ``The value relevance of digital marketing capabilities to firm performance,'' \emph{Journal of the Academy of Marketing Science}, vol.~50, no.~4, pp. 666--688, 2022.

\bibitem{cheriyan2018intelligent}
S.~Cheriyan, S.~Ibrahim, S.~Mohanan, and S.~Treesa, ``Intelligent sales prediction using machine learning techniques,'' in \emph{2018 International Conference on Computing, Electronics and Communications Engineering}.\hskip 1em plus 0.5em minus 0.4em\relax IEEE, 2018, pp. 53--58.

\bibitem{xia2020forexgboost}
Z.~Xia, S.~Xue, L.~Wu, J.~Sun, Y.~Chen, and R.~Zhang, ``Forexgboost: passenger car sales prediction based on xgboost,'' \emph{Distributed and Parallel Databases}, vol.~38, pp. 713--738, 2020.

\bibitem{geva2017using}
T.~Geva, G.~Oestreicher-Singer, N.~Efron, and Y.~Shimshoni, ``Using forum and search data for sales prediction of high-involvement projects,'' \emph{Mis Quarterly}, vol.~41, no.~1, pp. 65--82, 2017.

\bibitem{chen2011role}
Y.~Chen, S.~Fay, and Q.~Wang, ``The role of marketing in social media: How online consumer reviews evolve,'' \emph{Journal of interactive marketing}, vol.~25, no.~2, pp. 85--94, 2011.

\bibitem{landwehr2011gut}
J.~R. Landwehr, A.~A. Labroo, and A.~Herrmann, ``Gut liking for the ordinary: Incorporating design fluency improves automobile sales forecasts,'' \emph{Marketing Science}, vol.~30, no.~3, pp. 416--429, 2011.

\bibitem{sa2012multi}
A.~Sa-Ngasoongsong, S.~T. Bukkapatnam, J.~Kim, P.~S. Iyer, and R.~Suresh, ``Multi-step sales forecasting in automotive industry based on structural relationship identification,'' \emph{International Journal of Production Economics}, vol. 140, no.~2, pp. 875--887, 2012.

\bibitem{wang2018dynamic}
J.-N. Wang, J.~Du, Y.-L. Chiu, and J.~Li, ``Dynamic effects of customer experience levels on durable product satisfaction: Price and popularity moderation,'' \emph{Electronic Commerce Research and Applications}, vol.~28, pp. 16--29, 2018.

\bibitem{hulsmann2012general}
M.~H{\"u}lsmann, D.~Borscheid, C.~M. Friedrich, and D.~Reith, ``General sales forecast models for automobile markets and their analysis.'' \emph{Trans. Mach. Learn. Data Min.}, vol.~5, no.~2, pp. 65--86, 2012.

\bibitem{robinson1985sources}
W.~T. Robinson and C.~Fornell, ``Sources of market pioneer advantages in consumer goods industries,'' \emph{Journal of Marketing Research}, vol.~22, no.~3, pp. 305--317, 1985.

\bibitem{haubl1996cross}
G.~H{\"a}ubl, ``A cross-national investigation of the effects of country of origin and brand name on the evaluation of a new car,'' \emph{International Marketing Review}, vol.~13, no.~5, pp. 76--97, 1996.

\bibitem{urban1990prelaunch}
G.~L. Urban, J.~R. Hauser, and J.~H. Roberts, ``Prelaunch forecasting of new automobiles,'' \emph{Management Science}, vol.~36, no.~4, pp. 401--421, 1990.

\bibitem{assuncao1993rational}
J.~L. Assuncao and R.~J. Meyer, ``The rational effect of price promotions on sales and consumption,'' \emph{Management Science}, vol.~39, no.~5, pp. 517--535, 1993.

\bibitem{busse2010best}
M.~R. Busse, D.~I. Simester, and F.~Zettelmeyer, ``“the best price you'll ever get”: The 2005 employee discount pricing promotions in the us automobile industry,'' \emph{Marketing science}, vol.~29, no.~2, pp. 268--290, 2010.

\bibitem{hu2014ratings}
N.~Hu, N.~S. Koh, and S.~K. Reddy, ``Ratings lead you to the product, reviews help you clinch it? the mediating role of online review sentiments on product sales,'' \emph{Decision support systems}, vol.~57, pp. 42--53, 2014.

\bibitem{zahoor2023can}
A.~Zahoor, Y.~Yu, H.~Zhang, B.~Nihed, S.~Afrane, S.~Peng, A.~S{\'a}pi, C.~J. Lin, and G.~Mao, ``Can the new energy vehicles (nevs) and power battery industry help china to meet the carbon neutrality goal before 2060?'' \emph{Journal of Environmental Management}, vol. 336, p. 117663, 2023.

\bibitem{shilong2021machine}
Z.~Shilong \emph{et~al.}, ``Machine learning model for sales forecasting by using xgboost,'' in \emph{2021 IEEE International Conference on Consumer Electronics and Computer Engineering}.\hskip 1em plus 0.5em minus 0.4em\relax IEEE, 2021, pp. 480--483.

\bibitem{marcilio2020explanations}
W.~E. Marc{\'\i}lio and D.~M. Eler, ``From explanations to feature selection: assessing shap values as feature selection mechanism,'' in \emph{2020 33rd SIBGRAPI conference on Graphics, Patterns and Images}.\hskip 1em plus 0.5em minus 0.4em\relax Ieee, 2020, pp. 340--347.

\bibitem{baptista2022relation}
M.~L. Baptista, K.~Goebel, and E.~M. Henriques, ``Relation between prognostics predictor evaluation metrics and local interpretability shap values,'' \emph{Artificial Intelligence}, vol. 306, p. 103667, 2022.

\bibitem{chen2008online}
Y.~Chen and J.~Xie, ``Online consumer review: Word-of-mouth as a new element of marketing communication mix,'' \emph{Management science}, vol.~54, no.~3, pp. 477--491, 2008.

\bibitem{li2019effect}
X.~Li, C.~Wu, and F.~Mai, ``The effect of online reviews on product sales: A joint sentiment-topic analysis,'' \emph{Information and Management}, vol.~56, no.~2, pp. 172--184, 2019.

\end{thebibliography}

\end{document}